# Improvement of superconducting properties by chemical pressure effect in Eu-doped La$_{2-x}$Eu$_x$O$_2$Bi$_3$Ag$_{0.6}$Sn$_{0.4}$S$_6$

Rajveer Jha[1], Yosuke Goto[1], Ryuji Higashinaka[1], Akira Miura[2], C. Moriyoshi[3], Yoshihiro Kuroiwa[3] and Yoshikazu Mizuguchi[1]

[1] Department of Physics, Tokyo Metropolitan University, 1-1 Minami-Osawa, Hachioji, Tokyo 192-0397,Japan
[2] Faculty of Engineering, Hokkaido University, Kita-13, Nishi-8, Kita-ku, Sapporo 060-8628, Japan
[3] Department of Physical Science, Hiroshima University, 1-3-1 Kagamiyama, Higashihiroshima, Hiroshima 739-8526, Japan.

E-mail: rajveerjha@gmail.com

**Abstract**

We have investigated the substitution effect of Eu on the superconductivity in La$_{2-x}$Eu$_x$O$_2$Bi$_3$Ag$_{0.6}$Sn$_{0.4}$S$_6$. Recently, we reported an observation of superconductivity at 0.5 K in a layered oxychalcogenide La$_2$O$_2$Bi$_3$AgS$_6$. The Sn doping at the Ag site was found to raise the superconducting transition temperature, $T_c$ to 2.5 K in La$_2$O$_2$Bi$_3$Ag$_{0.6}$Sn$_{0.4}$S$_6$. To further improve the superconducting properties, we have partially substituted Eu for the La site to increase the chemical pressure in La$_{2-x}$Eu$_x$O$_2$Bi$_3$Ag$_{0.6}$Sn$_{0.4}$S$_6$ ($x$ = 0.1–0.6). With the increase in Eu concentration, $x$, the lattice constant $a$ was found to shrink, while the lattice constant $c$ was marginally shortened, which suggests that the chemical pressure induced by the Eu doping is uniaxial along the $a$-axis. $T_c$ was observed to increase with increasing $x$ up to $x$ = 0.4, further decreasing for higher Eu concentrations of $x$ = 0.5 and 0.6. From the magnetic susceptibility and resistivity measurements, the bulk nature of superconductivity has been observed for $x$ = 0.1–0.5 with $T_c$ = 2.5–4.0 K, respectively. The upper critical field ($B_{c2}$) was noted to be 3.5 T for $x$ = 0.4, which also has the highest $T_c$.

## 1. Introduction

The recent discovery of a family of layered superconductors based on $BiS_2$, $Bi_4O_4S_3$, $REO_{1-x}F_xBiS_2$ (RE = La, Ce, Nd, Yb, Pr), and $Sr_{1-x}La_xFBiS_2$, having superconducting transition temperatures ($T_c$) in the range of of 2–10.5 K [1-16] has provided a new platform for designing and investigating layered superconductors. The basic crystal structures of the $BiS_2$-based compounds involve alternate stacking of two $BiS_2$ conducting layers ($BiS_2$ bilayer) and a blocking layer like that of $La_2O_2$. Due to the stacked layered structure, various kinds of $BiS_2$-based superconductors have been designed such as high-$T_c$ cuprates and Fe-based superconductors[18,19] by changing theconstituent elements and blocking layer structures [17]. In addition, the crystal structure of the $BiS_2$-based compounds is quite interesting because of the presence of van der Waals gaps in between two superconducting $BiS_2$ layers. This feature allowed us to design a thick superconducting layer in $La_2O_2M_4S_6$ ($M$: Pb, Ag, Bi, Sn) by inserting NaCl-type $MS$ layers into the van der Waals gaps [20-22]. Recently, we reported an observation of superconductivity at 0.5 K in a layered oxychalcogenide $La_2O_2Bi_3AgS_6$ [23], which motivated us to further study substitution effects on the superconductivity of the $La_2O_2M_4S_6$-type systems. To this end, we had previously doped Sn at the Ag site in $La_2O_2Bi_3Ag_{0.6}Sn_{0.4}S_6$ and found that $T_c$ increased up to 2.5 K [24]. Although the superconducting properties had been improved to a certain extent, the chemical pressure obtained was not sufficient with the Sn doping. In-plane chemical pressure is one of the key parameters to induce bulk superconductivity in the $BiS_2$-based systems because it suppresses in-plane structural disorder due to the presence of Bi lone pairs [25-27]. Using this concept, we had achieved bulk superconductivity at 3.5 K for $La_2O_2Bi_3Ag_{0.6}Sn_{0.4}S_{5.7}Se_{0.3}$ via Se partial substitution [24]. However, the solubility limit of Se for the S site was very limited (around 5%-Se). Therefore, we have explored a different site substitution to further investigate the in-plane chemical pressure effect in the $La_2O_2M_4S_6$ system.

In this study, we have shown the emergence of bulk superconductivity in $La_{2-x}Eu_xO_2Bi_3Ag_{0.6}Sn_{0.4}S_6$ via the La-site substitution by Eu. The highest superconducting transition temperature, $T_c$, of 4 K was observed for $x = 0.4$ (20%-Eu) sample. This is validated by the variation of magnetization measurements of the sample with temperature, which shows evidence of bulk superconductivity with a shielding volume fraction of more than 75%. Moreover, we have systematically doped Eu ($x = 0.1$ to 0.6) at the La site and observed the shrinking of lattice parameters $a$ and $c$ with increasing Eu concentration, $x$, in $La_{2-x}Eu_xO_2Bi_3Ag_{0.6}Sn_{0.4}S_6$, which suggests an in-plane chemical pressure effect due to the Eu doping. Successively, the superconducting properties were observed to improve with increasing Eu concentration in $La_{2-x}Eu_xO_2Bi_3Ag_{0.6}Sn_{0.4}S_6$.

## 2. Experimental details

The polycrystalline samples of $La_{2-x}Eu_xO_2Bi_3Ag_{0.6}Sn_{0.4}S_6$ were prepared by a solid-state reaction method. Powders of EuS (99.99%), $Bi_2O_3$ (99.9%), $La_2S_3$ (99.9%), Sn (99.99%), AgO (99.9%), S (99.99%), and grains of Bi (99.999%) with a nominal composition of $La_{2-x}Eu_xO_2Bi_3Ag_{0.6}Sn_{0.4}S_6$ ($x = 0, 0.1, 0.2, 0.3, 0.4, 0.5,$ and 0.6) were mixed using a pestle and a mortar. The mixed powder was pelletized, sealed in an evacuated quartz tube, and heated at 720 °C for 15 h. To homogenize the sample, the product was re-grounded, pelletized, and heated at 720 °C for 15 h. The phase purity of the prepared samples and the optimal annealing conditions were examined using laboratory X-ray diffraction (XRD) by the θ-2θ method with Cu-K$_α$ radiation. The synchrotron XRD experiment was performed at BL02B2, SPring-8 under the proposal of [2019A1101]. The wavelength of the synchrotron X-ray used in this study was 0.49581 Å. The crystal structure parameters were refined using the Rietveld method with RIETAN-FP [28]. A schematic image of the crystal structure was drawn using VESTA (see the inset of Fig. 1) [29]. The actual compositions of the synthesized samples were investigated by energy-dispersive X-ray spectroscopy (EDX) with TM-3030 (Hitachi). Electrical resistivity for temperatures above 2 K was measured by the four-probe technique using the Physical Property measurement system (PPMS: Quantum Design). The temperature dependence of magnetic susceptibility χ (T) was measured using a superconducting quantum interference device (SQUID) magnetometer (MPMS-3, Quantum Design).

## 3. Results and discussion

Figure 1(a) shows the powder XRD pattern of $La_{2-x}Eu_xO_2Bi_3Ag_{0.6}Sn_{0.4}S_6$ ($x$ = 0–0.6) at room temperature. The $La_{2-x}Eu_xO_2Bi_3Ag_{0.6}Sn_{0.4}S_6$ samples obtained are crystallized in the tetragonal structure with the space group of P4/nmm. Essentially, the $La_{2-x}Eu_xO_2Bi_3Ag_{0.6}Sn_{0.4}S_6$ is composed of stacked [$M_4S_6$] layers and fluorite-type [$La_2O_2$] layers. In this study, Eu was doped in place of the La site in the fluorite-type [$La_2O_2$] layer. We observed a minor impurity phase of $Eu_2Sn_2O$ denoted by an asterisk in Fig. 1(a) along with the small impurity peaks of $Bi_2S_3$ for high Eu concentration ($x \geq 0.4$) samples marked by the "$" symbol. The shift of 103 peak position is shown in Fig. 1(b). As the Eu concentration increases from $x$ = 0 to x = 0.4, the 103 peak shifts toward the lower angle side, while for $x$ > 0.4, the 103 peak shifts slightly toward the higher angle side. The schematic image of the unit cell for $La_{1.6}Eu_{0.4}O_2Bi_3Ag_{0.6}Sn_{0.4}S_6$ is shown in Fig. 1(c), where the MS layers of $(BiAg_{0.6}Sn_{0.4})S$ and $La_{2-x}Eu_xO(Bi_{0.9}Ag_{0.06}Sn_{0.04})S_2$ are alternately stacked. The evolution of the unit cell parameters with the amount of Eu doping, obtained from Rietveld fits to the powder patterns, are shown in Fig.1 (d). The estimated lattice parameters are a = 4.064(1) Å and c = 19.447(1) Å for x = 0, and a = 4.051(1) Å and c = 19. 43(1) Å for x = 0.4. The lattice parameter, c, first decreases for $x$ = 0–0.4 and then increases for higher values of $x$. The lattice parameter, a, decreases sharply for $x$ = 0–0.4, which suggests compression in the a-b plane. This shrinkage caused by the Eu doping induces in-plane chemical pressure in the superconducting layer of $La_{2-x}Eu_xO_2Bi_3Ag_{0.6}Sn_{0.4}S_6$. For samples with a higher Eu concentration, i.e., $x$ = 0.5 and 0.6, the lattice parameters, $a$ and $c$, slightly increase, which suggests that the solubility limit of the Eu doping in $La_{2-x}Eu_xO_2Bi_3Ag_{0.6}Sn_{0.4}S_6$ is more or less $x$ ~ 0.4. The Eu concentrations in the $La_{2-x}Eu_xO_2Bi_3Ag_{0.6}Sn_{0.4}S_6$ ($x$ = 0.1-0.6) samples were examined using EDX. The nominal $x$ and the obtained values of $x_{EDX}$ are shown in Fig.1 (g). We can infer from Fig.1 (g) that $x_{EDX}$ is slightly deviating from the red dotted line with increasing x, suggesting that the actual concentration of Eu is lower than the nominal concentration.

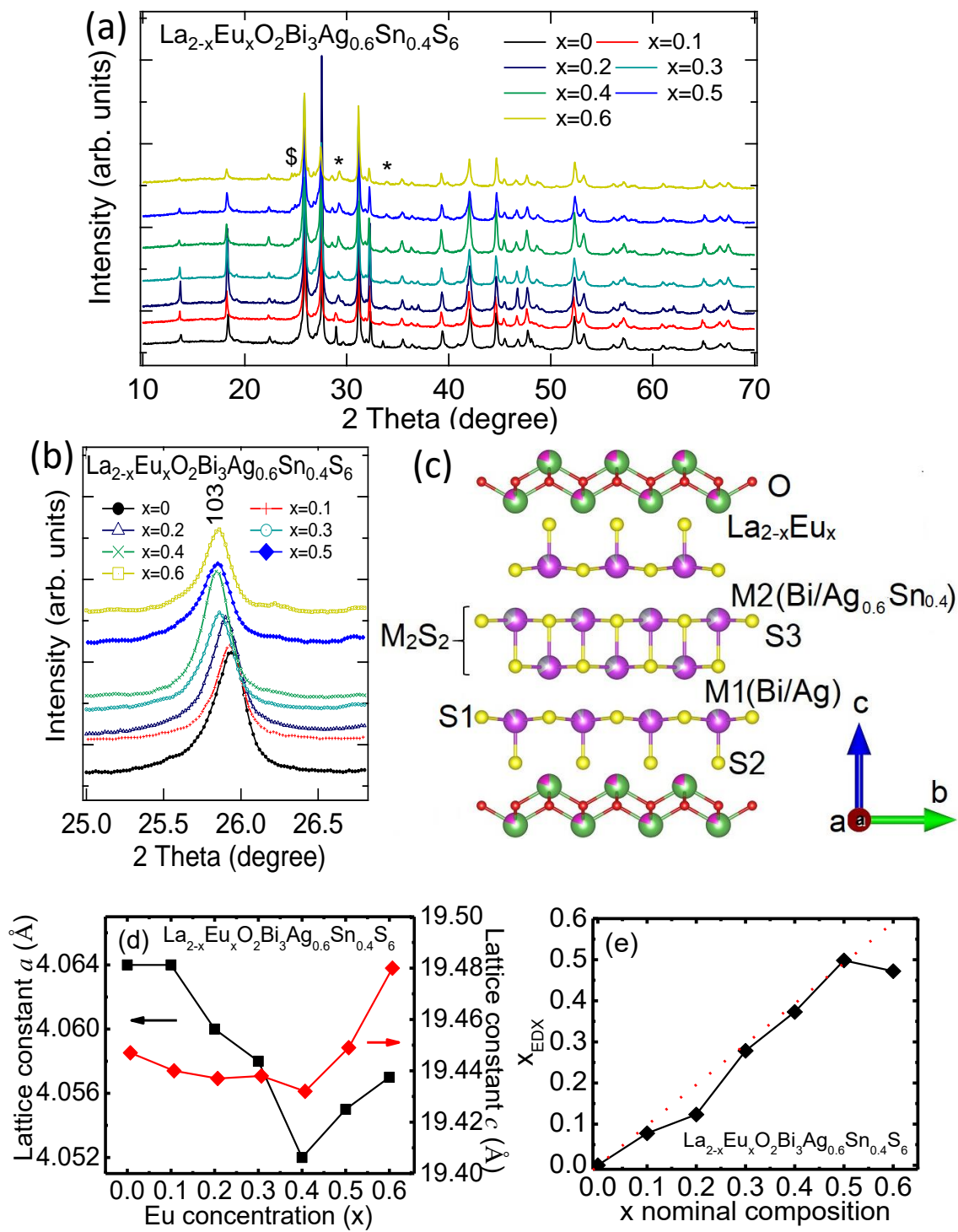

**Fig. 1:** (color online) (a) The room temperature XRD pattern of $La_{2-x}Eu_xO_2Bi_3Ag_{0.6}Sn_{0.4}S_6$ ($x = 0$-$0.6$) compounds. (b) The XRD pattern near to 103 (Miller indices) peak of tetragonal phase of $La_2O_2Bi_3AgS_6$. (c) The schematic unit cell of $La_{1.6}Eu_{0.4}O_2Bi_3Ag_{0.6}Sn_{0.4}S_6$ compound. (d) Lattice parameters $a$ and $c$ obtained by a refinement using the Rietveld method with RIETAN-FP [23]. (e) The nominal composition of Eu in $La_{2-x}Eu_xO_2Bi_3Ag_{0.6}Sn_{0.4}S_6$ vs Eu concentration obtained by EDX.

Figure 2 shows the synchrotron XRD (SXRD) pattern and the Rietveld refinement result for the La$_{1.6}$Eu$_{0.4}$O$_2$Bi$_3$Ag$_{0.6}$Sn$_{0.4}$S$_6$ sample. The SXRD pattern was refined using multi-phase analysis with the main phase identified with the tetragonal P4/nmm space group of La$_2$O$_2$Bi$_3$AgS$_6$-type and the impurity phases identified with Bi$_2$S$_3$(5%) and Eu$_2$Sn$_2$O$_7$(4.4%). The refined lattice constants are $a = 4.04984(5)$Å and $c = 19.42801(4)$Å. The reliability factor $R_{wp}$ is 11.6% and the refined lattice parameters are close to those determined by laboratory XRD data analysis.

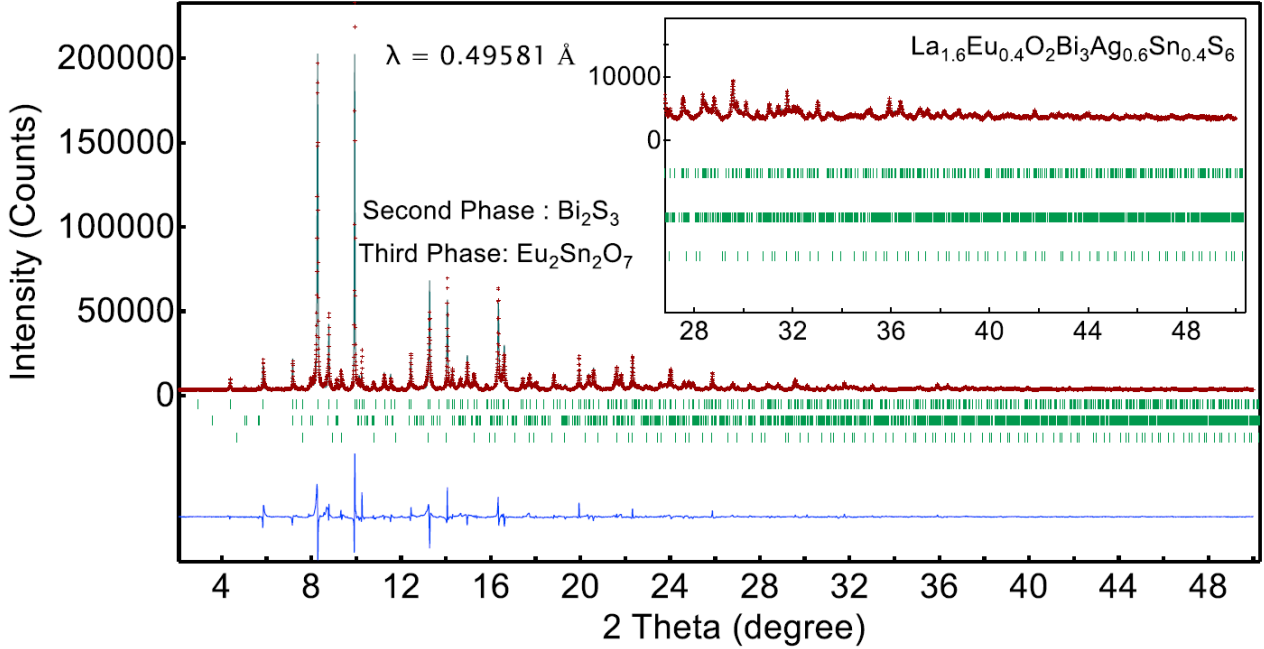

**Fig. 2:** The synchrotron XRD pattern and the Rietveld analysis for the La$_{1.6}$Eu$_{0.4}$O$_2$Bi$_3$Ag$_{0.6}$Sn$_{0.4}$S$_6$ sample is shown. The Rietveld analysis was performed by multi-phase refinement with the Bi$_2$S$_3$ and Eu$_2$Sn$_2$O$_7$ impurity phases. The inset shows an expanded profile at higher angles.

Figure 3 (a–g) shows the temperature dependence of magnetic susceptibility $\chi(T)$ for La$_{2-x}$Eu$_x$O$_2$Bi$_3$Ag$_{0.6}$Sn$_{0.4}$S$_6$ ($x =0$-0.6) under an applied magnetic field ($B$) strength of 1 mT. The magnetic susceptibility measurements were performed using two protocols: zero field-cooled (ZFC) and field-cooled (FC). A diamagnetic signal was observed below 2.2 K, 2.8 K, 3.3 K, 3.4 K, 4.0 K, 4.1 K, and 4.1 K for $x = $ 0.0, 0.1, 0.2, 0.3, 0.4, 0.5, and 0.6, respectively. A large diamagnetic signal was observed below 4.0 K in the ZFC curve for $x = 0.4$. The unsaturated value of $4\pi\chi$ (ZFC) at 2.0 K with the shielding volume fraction at the corresponding temperature being above 75% indicates that the bulk superconductivity has been induced by Eu substitution in La$_{2-x}$Eu$_x$O$_2$Bi$_3$Ag$_{0.6}$Sn$_{0.4}$S$_6$. The nominal $x$ dependence on the shielding volume fraction is shown in Fig. 3(h). The volume fraction increases with increasing Eu concentration up to $x = 0.4$ and further decreases for $x = 0.5$ and 0.6. On the other hand, $T_c$ obtained for the latter samples is slightly higher than that for the former samples. The decreased volume fraction for higher $x$ can be understood by the large amount of impurity phases in the samples

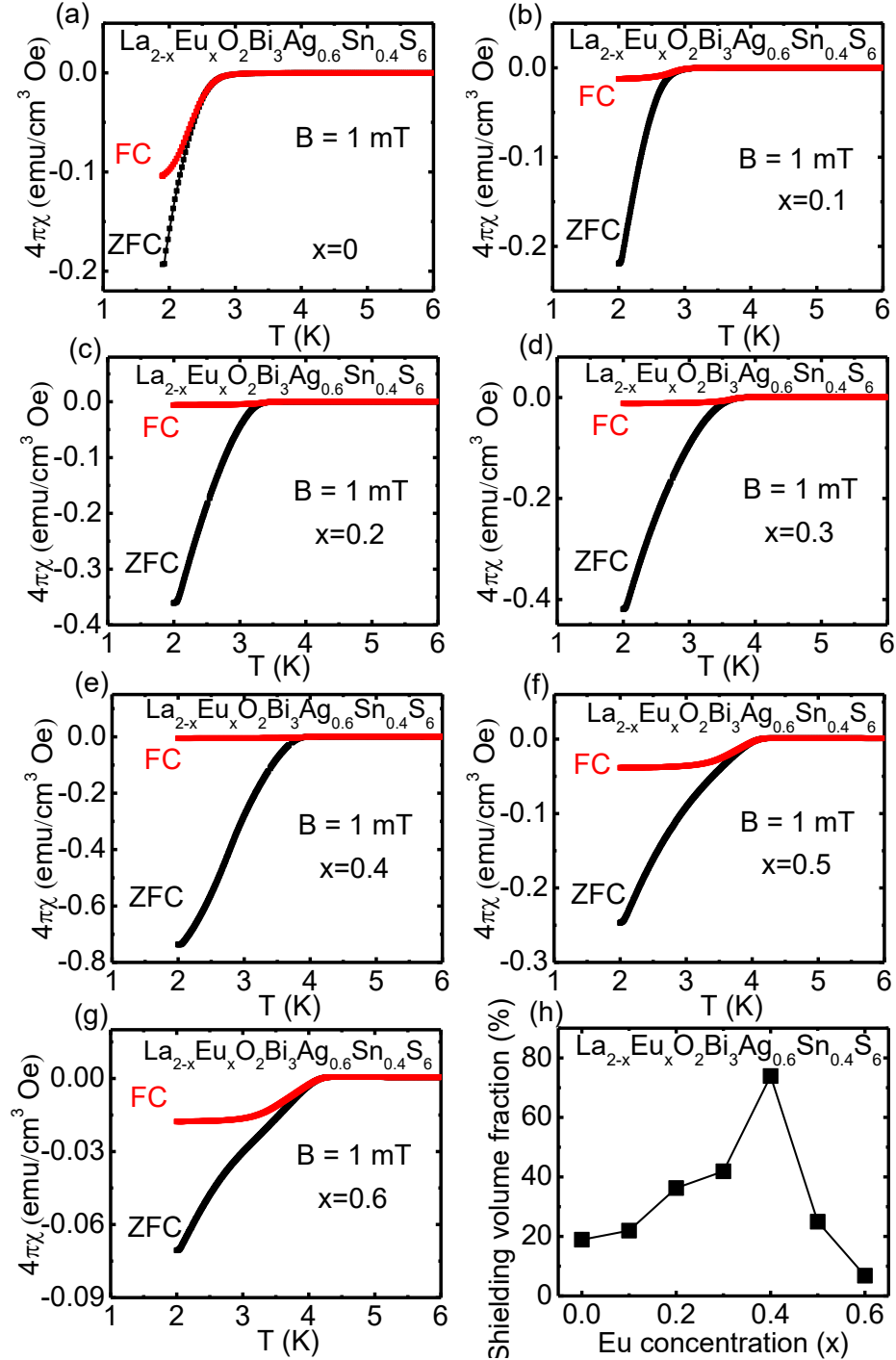

**Fig. 3:** (a-g) Temperature (T) dependence of magnetic susceptibility ($\chi$) for La$_{2-x}$Eu$_x$O$_2$Bi$_3$Ag$_{0.6}$Sn$_{0.4}$S$_6$ ($x$ = 0–0.6) compounds measured in the ZFC and the FC protocol for an applied magnetic field of 1 mT. (h) Eu concentration (x) dependence on the shielding volume fraction of La$_{2-x}$Eu$_x$O$_2$Bi$_3$Ag$_{0.6}$Sn$_{0.4}$S$_6$ ($x$ = 0-0.6).

Figure 4 (a) shows the temperature dependence of $\chi$ for $La_{2-x}Eu_xO_2Bi_3Ag_{0.6}Sn_{0.4}S_6$ ($x = 0.1–0.6$) under a magnetic field, $B = 1$ T, for temperatures in the range 300 to 2 K. All the Eu-doped samples show the Curie–Weiss behavior. We analyzed the $\chi$-$T$ data between 5 and 70 K using the Curie–Weiss law [30]. The log-log plot for the $(\chi - \chi_0)^{-1}$ vs $T$ in Fig. 4(b) shows a linear behavior for temperatures between 5 K and 70 K. We selected data in this range for fitting with the Curie–Weiss law provided as follows,

$$\chi = \chi_0 + \frac{N_a \mu_{eff}^2 \mu_B^2}{3 k_B (T - \theta)} = \chi_0 + \frac{C}{(T - T_{CW})}, \quad \ldots (1)$$

Here, $\chi_0$ is the temperature independent term of the Pauli contribution, $N_a$ is the Avogadro number, $\mu_B$ is the Bohr magneton, $k_B$ is the Boltzmann constant, $T$ is the temperature, $T_{CW}$ is the Curie–Weiss temperature, $\mu_{eff}$ is the effective magnetic moment, and $C = N_a \mu_{eff}^2 \mu_B^2 / 3 k_B$ is the Curie constant. The Curie–Weiss law in the form given in equation (1) was fitted to the selected data to determine the effective magnetic moments ($\mu_{eff} = \mu_{Eu}$) and the dominant magnetic ordering (sign of $T_{CW}$). Fig. 4 (c, d) shows the estimated effective magnetic moments and the Weiss temperature for $La_{2-x}Eu_xO_2Bi_3Ag_{0.6}Sn_{0.4}S_6$ ($x = 0.1–0.6$). The estimated effective magnetic moments are $\mu_{Eu} = 0.483 \mu_B$/Eu for $x = 0.1$, $\mu_{Eu} = 0.699 \mu_B$/Eu for $x = 0.2$, $\mu_{Eu} = 0.770 \mu_B$/Eu for $x = 0.3$, $\mu_{Eu} = 0.852 \mu_B$/Eu for $x = 0.4$, $\mu_{Eu} = 0.791 \mu_B$/Eu for $x = 0.5$ and $\mu_{Eu} = 0.796 \mu_B$/Eu for $x = 0.6$. Their corresponding Curie-Weiss temperature ($T_{CW}$) are -0.4, -0.6, -0.8, -1.1, -1.3, and -1.1 K. The obtained values from the fitting of the Curie–Weiss law are summarized in Table 1. The trend observed in

Fig. 4 (c, d) depicts an increase in $\mu_{eff}$ and the magnitude of $T_{CW}$ with the increase in Eu concentration. This suggests that the antiferromagnetic interaction is enhanced by the Eu substitution. The negative values of $T_{CW}$ for large concentration of $La_{2-x}Eu_xO_2Bi_3Ag_{0.6}Sn_{0.4}S_6$ ($x = 0.1–0.6$) suggest a weak antiferromagnetic interaction of $Eu^{2+}$ ions in the $La_{2-x}Eu_xO_2Bi_3Ag_{0.6}Sn_{0.4}S_6$ ($x = 0.1–0.6$) samples. However, an antiferromagnetic transition was not observed in the magnetic susceptibility measurements for temperatures above 2 K for all the samples.

Thus, the calculation of the effective magnetic moment is straightforward. The calculated values are smaller than the effective magnetic moment of $Eu^{2+}$, i.e., 7.8 $\mu_B$. Our analysis suggests that although the doped Eu has a valence state close to $Eu^{3+}$ in the $La_{2-x}Eu_xO_2Bi_3Ag_{0.6}Sn_{0.4}S_6$ ($x = 0.1-0.6$) system, the valence state could contain a very small amount of $Eu^{2+}$

**Table 1**: The values of paramagnetic Susceptibility $\chi_0$ (emu/Eu mol-Oe), Curie constants $C$ (emu.K/ Eu mol-Oe), effective moment $\mu_{eff}$ ($\mu_B$/Eu) and Curie–Weiss temperatures $T_{CW}$ (K) for $La_{2-x}Eu_xO_2Bi_3Ag_{0.6}Sn_{0.4}S_6$ obtained from fitting Eq.(1) to the data.

| $x$ | $\chi_0$ (emu/Eu mol-Oe) | $C$ (emu.K/Eu mol-Oe) | $\mu_{eff}$ ($\mu_B$/Eu) | $T_{CW}$ (K) |
|---|---|---|---|---|
| 0.1 | 0.0063 | 0.0305 | 0.493 | 0.398 |
| 0.2 | 0.0071 | 0.0612 | 0.699 | 0.625 |
| 0.3 | 0.0057 | 0.0742 | 0.770 | 0.810 |
| 0.4 | 0.0067 | 0.0908 | 0.852 | -1.127 |
| 0.5 | 0.0066 | 0.0784 | 0.791 | -1.287 |
| 0.6 | 0.0063 | 0.0793 | 0.796 | -1.145 |

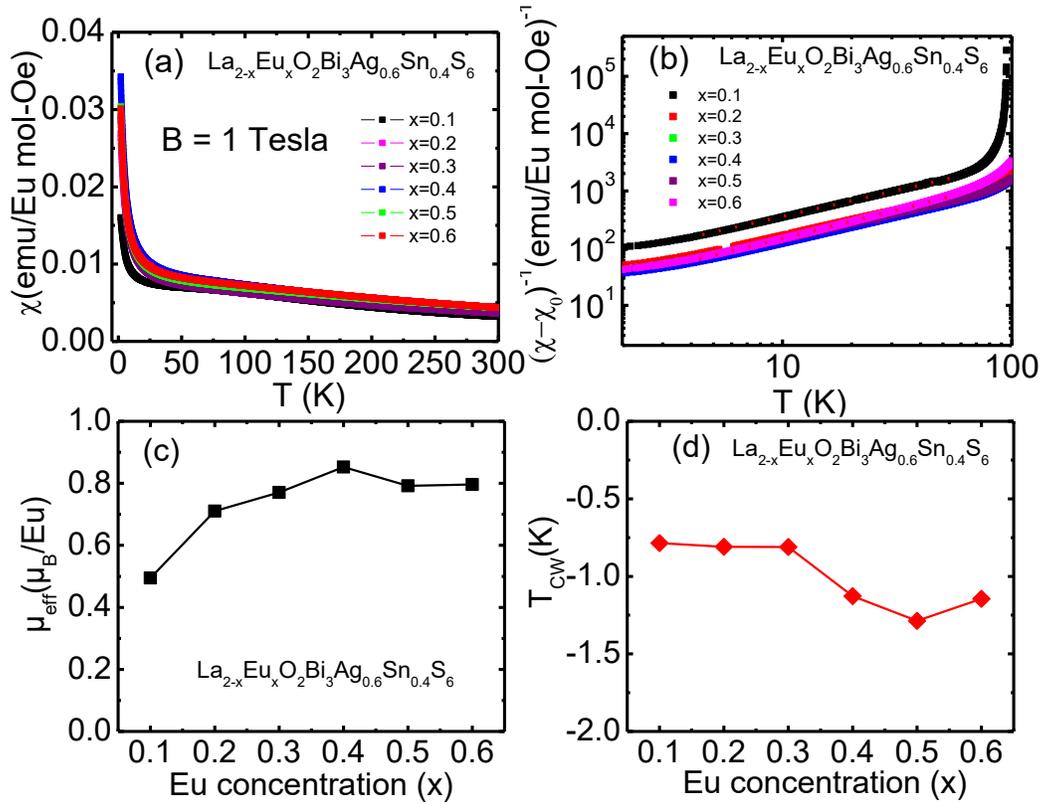

**Fig. 4:** (a) Temperature (T) variation of magnetic susceptibility ($\chi$) for the $La_{2-x}Eu_xO_2Bi_3Ag_{0.6}Sn_{0.4}S_6$ ($x$ = 0.1-0.6) compounds measured in the ZFC protocol from temperatures 300–2 K for an applied magnetic field of 1 T. (b) The ($\chi$- $\chi_0$)$^{-1}$ vs T for $La_{2-x}Eu_xO_2Bi_3Ag_{0.6}Sn_{0.4}S_6$ ($x$ = 0-0.6) with the red dotted line representing a linear variation for reference. (c, d) $x$ dependence of the effective moment and the Curie–Weiss temperature, respectively, for $La_{2-x}Eu_xO_2Bi_3Ag_{0.6}Sn_{0.4}S_6$ ($x$ = 0.1-0.6) obtained using the Curie–Weiss law.

Figure 5(a) displays the temperature dependence of electrical resistivity, $\rho(T)$, from 300 to 1.5 K for $La_{2-x}Eu_xO_2Bi_3Ag_{0.6}Sn_{0.4}S_6$ ($x$ = 0–0.6). The electrical resistivity at 300 K decreases moderately with increasing Eu concentration, x, up to $x$ = 0.4 and then increases again for $x$ = 0.5 and 0.6 samples. The normal state electrical resistivity of the Eu-doped compounds also shows a remarkable temperature dependence. For example, in the pure sample i.e., $x$ = 0, the resistivity rises for temperatures below 100 K and saturates near $T_c$. A semiconducting-type behavior in the normal state resistivity has been observed above $T_c$, which has been observed to undergo suppression with increasing Eu concentration in $La_{2-x}Eu_xO_2Bi_3Ag_{0.6}Sn_{0.4}S_6$. The latter can be understood by the in-plane chemical pressure effect as observed in $BiS_2$-based systems [17] and seems to be linked to the improved superconducting properties.

Figure 5(b) shows the zoomed view of Fig. 5(a) near the superconducting transition region. $T_c$ shifts monotonically towards the higher temperature side with increasing Eu concentration in $La_{2-x}Eu_xO_2Bi_3Ag_{0.6}Sn_{0.4}S_6$. The highest $T_c$ of 4.0 K was achieved for $x$ = 0.4 after which it starts decreasing for $x$ = 0.5 and 0.6. The lowering in $T_c$ for higher Eu-doped samples in the $\rho(T)$ data suggests that the presence of large amounts of impurity phases affects the resistive transition for $x$ = 0.5 and 0.6.

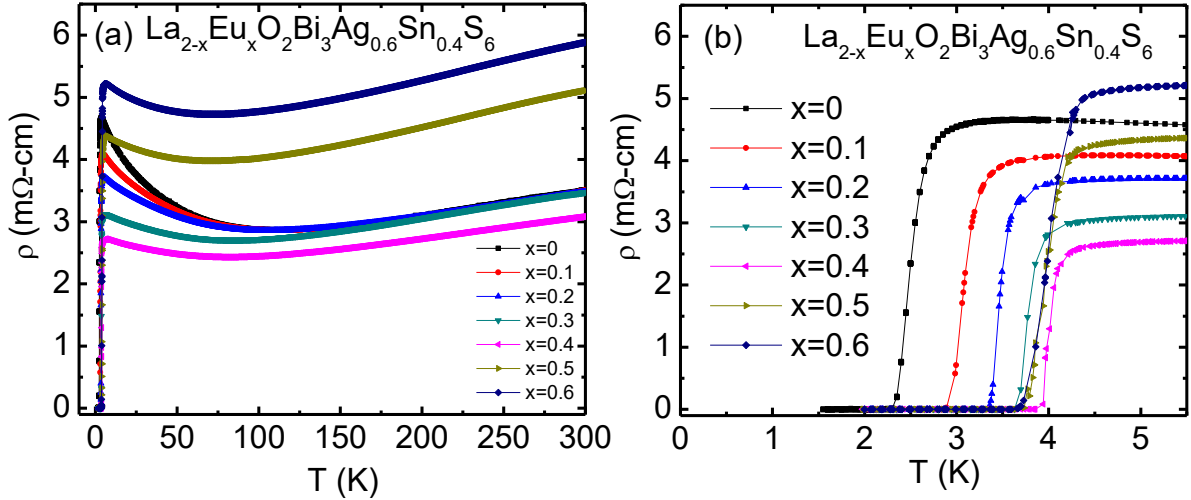

**Fig. 5:** (color online) (a) Temperature dependence of electrical resistivity measured from 300 K-1.5 K for the $La_{2-x}Eu_xO_2Bi_3Ag_{0.6}Sn_{0.4}S_6$ ($x$ = 0–0.6) compounds. (b) The $\rho(T)$ curve in the temperature range 5.5–1.5 K

Figure 6 (a-g) shows the temperature dependence of electrical resistivity under various magnetic fields, $B$, for $La_{2-x}Eu_xO_2Bi_3Ag_{0.6}Sn_{0.4}S_6$ ($x$ = 0–0.6). $T_c$ shifts towards the lower temperature side with increasing magnetic field. We evaluated the beginning of the drop in the resistivity due to the superconductivity at a higher magnetic field strength, $B$ = 3 T, for $x$ = 0.4 ($La_{2-x}Eu_xO_2Bi_3Ag_{0.6}Sn_{0.4}S_6$). $T_c$ onset shifts slowly with increasing magnetic field as compared to $T_c^{zero}$, which suggests the robustness of superconductivity against the magnetic field in the Eu-doped $La_{2-x}Eu_xO_2Bi_3Ag_{0.6}Sn_{0.4}S_6$ system. The upper critical field ($B_{c2}$) verses temperature phase diagrams are shown in Fig. 6(f). $B_{c2}$ are estimated at temperatures where the normal state resistivity ($\rho_n$) drops to 90% for various applied magnetic fields. We have estimated $B_{c2}(0)$ by the conventional one-band Werthamer–Helfand–Hohenberg (WHH) equation [31], i.e., $B_{c2}(0) = -0.693T_c(dB_{c2}/dT)_{T=T_c}$. The values of $B_{c2}(0)$ obtained from the WHH method are 1.8, 1.87, 2.425, 2.75, 3.45, 2.14, and 1.87 T for $x$ = 0, 0.1, 0.2, 0.3, 0.4, 0.5, and 0.6, respectively.

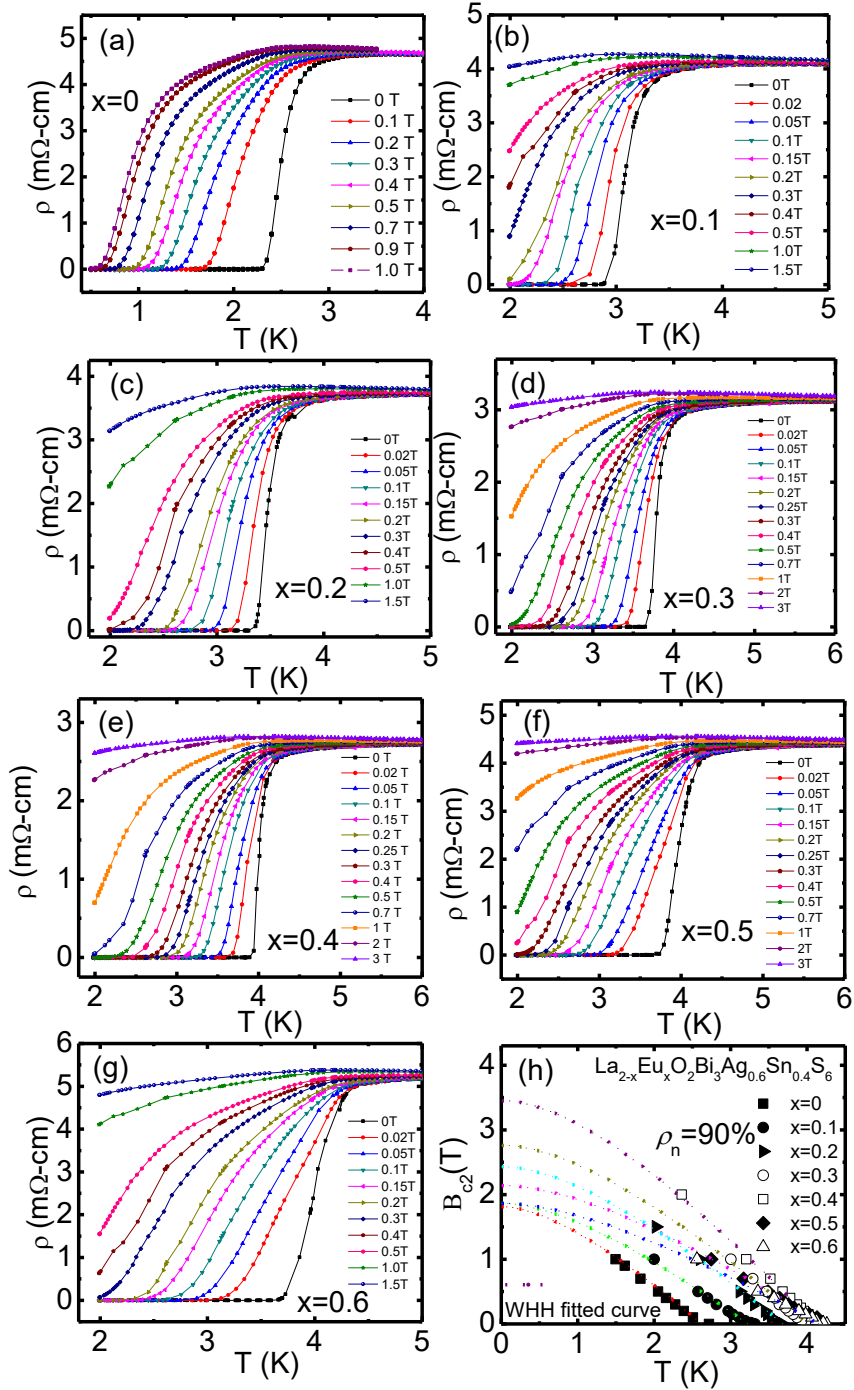

**Fig. 6:** (color online) (a) The temperature dependence of electrical resistivity from 3.0–0.5 K for the $La_{2-x}Eu_xO_2Bi_3Ag_{0.6}Sn_{0.4}S_6$ ($x$ = 0) compound for various magnetic fields. (b–g) The temperature dependence of electrical resistivity from 6.0-2.0 K for the La$_{2-x}$Eu$_x$O$_2$Bi$_3$Ag$_{0.6}$Sn$_{0.4}$S$_6$ ($x$=0.1–0.6) compounds for different magnetic fields. (h) The temperature dependence of $B_{c2}$(T) for La$_{2-x}$Eu$_x$O$_2$Bi$_3$Ag$_{0.6}$Sn$_{0.4}$S$_6$($x$=0–0.6). Experimental data shown in symbols, are fitted with the WHH theory (dotted lines).

To estimate the change in carrier concentration, Seebeck coefficients for La$_{2-x}$Eu$_x$O$_2$Bi$_3$Ag$_{0.6}$Sn$_{0.4}$S$_6$ ($x$ = 0–0.6) were measured by a four-probe method at room temperature as depicted in Fig. 7. We observed no clear change in the carrier concentration by the Eu substitution. The magnitude of Seebeck coefficient ($S$) increases abruptly for $x$ = 0.5 and 0.6, which suggests that the electron concentration decreases for higher Eu concentrated samples which in turn might be due to the increased amount of impurity phase for higher $x$.

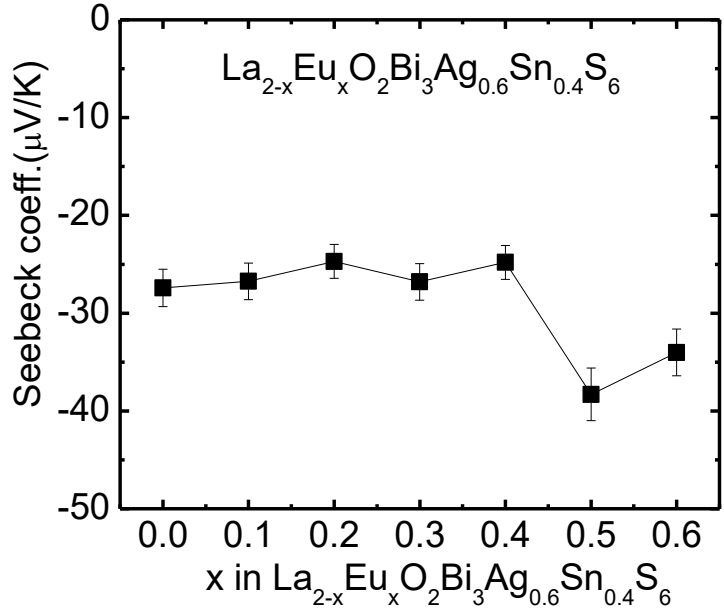

**Fig. 7:** The x dependence of Seebeck coefficient for La$_{2-x}$Eu$_x$O$_2$Bi$_3$Ag$_{0.6}$Sn$_{0.4}$S$_6$ compounds at room temperature.

Figure 8 exhibits the $T_c$ vs $x$ phase diagram for La$_{2-x}$Eu$_x$O$_2$Bi$_3$Ag$_{0.6}$Sn$_{0.4}$S$_6$. The superconducting transition temperature, $T_c^{zero}$, obtained from $\rho(T)$ measurements and $T_c$ obtained from the $\chi$-$T$ curve are plotted as functions of Eu concentration, $x$, in Fig. 8. $T_c$ gradually increases with $x$ in La$_{2-x}$Eu$_x$O$_2$Bi$_3$Ag$_{0.6}$Sn$_{0.4}$S$_6$. The highest $T_c^{zero}$ of 4.0 K was achieved for $x$ = 0.4 from both the measurements, $T_c^{zero}$ decreases for $x$ = 0.5 and 0.6, while the magnetic $T_c$ in diamagnetic transition increases slightly for the $x$ = 0.5 and 0.6 compounds. The latter can be understood from the presence of small particles with high Eu concentration in the examined sample. However, as mentioned above, the samples with $x$ = 0.5 and 0.6 contains many impurity phases due to the solubility limit of Eu for the La site in this system. In addition, the shielding volume fraction for those samples were smaller than that for $x$ = 0.4. We assume that the difference in $T_c$ between the resistivity and the magnetic measurements occurred due to the presence of higher-$T_c$ particles in $x$ = 0.5 and 0.6.

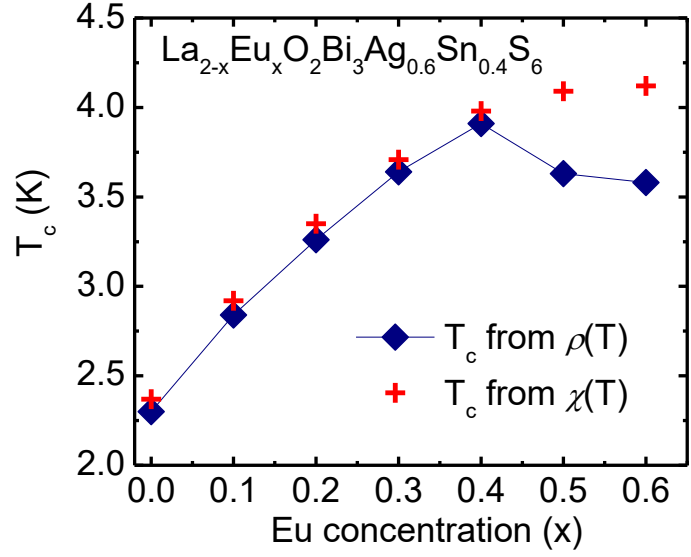

**Fig. 8:** The $x$ dependence of $T_c$ for $La_{2-x}Eu_xO_2Bi_3Ag_{0.6}Sn_{0.4}S_6$ compounds obtained from $\rho(T)$ and $\chi(T)$ data.

In summary, we have investigated the substitution effect of Eu for the La site in $La_{2-x}Eu_xO_2Bi_3Ag_{0.6}Sn_{0.4}S_6$. $T_c$ was observed to increase with increasing Eu concentration. The highest $T_c^{zero}$ of 4.0 K was observed for $x = 0.4$. Bulk nature of superconductivity has been confirmed through $\chi$-$T$ measurements with $T_c = 4.0$ K and the corresponding shielding volume fraction exceeding 75% for $x = 0.4$. The structural analysis suggests that the solubility limit of Eu in $La_{2-x}Eu_xO_2Bi_3Ag_{0.6}Sn_{0.4}S_6$ system is ~20% ($x$~0.4), representative of high superconducting properties. The estimated upper critical field $B_{c2}(0)$ for $La_{2-x}Eu_xO_2Bi_3Ag_{0.6}Sn_{0.4}S_6$ ($x = 0$–0.6) are 1.8, 1.87, 2.425, 2.75, 3.45, 2.14 and 1.87 T for $x = 0, 0.1, 0.2, 0.3, 0.4, 0.5$ and 0.6, respectively.

Our present results will motivate further studies of $BiS_2$-based superconductors with a ticker conducting layer, especially the effects of tuning chemical pressure and spacer layers in $La_2O_2M_4S_6$-type compounds.

Recently we found a paper reporting La-substitution effect by Nd and Pr [32]. Those substitution can also be regarded as chemical pressure.

### Acknowledgements

We gratefully appreciate O. Miura of Tokyo Metropolitan University for fruitful discussions. This work was financially supported by grants in Aid for Scientific Research (KAKENHI) (Grant Nos. 15H05886, 15H05884, 16H04493, 17K19058, 16K05454, and 15H03693)